\documentclass[doublecol,amsmath,amssymb]{epl2} 
\usepackage{graphicx}
\usepackage{amsmath}
\usepackage{amssymb}
\usepackage{bm}

\title{ {Emergence of collective intonation in the musical performance of crowds}}
\shorttitle{Emergence of collective intonation in crowds} 

\author{Lucas Lacasa}

\institute{                    
  School of Mathematical Sciences, Queen Mary University of London, Mile End Road E14NS London (UK)
}
\pacs{43.66.Lj}{Perceptual effects of sound}
\pacs{89.75.-k}{Complex Systems}

\abstract{
The average individual is typically a mediocre singer, with a rather restricted capacity to sing a melody in tune. Yet when many singers are assembled to perform collectively, the resulting melody of the crowd is suddenly perceived by an external listener as perfectly tuned -as if it was actually a choral performance- even if each individual singer is out of tune. This collective phenomenon is an example of a wisdom of crowds effect that can be routinely observed in music concerts or other social events, when a group of people spontaneously sings at unison. In this paper we rely on the psychoacoustic properties of pitch and provide a simple mechanistic explanation for the onset of this emergent behavior.}

\begin{document}

\maketitle

The wisdom of crowds \cite{book} is a popular concept englobing several examples of collective intelligence, that emerges where the collective response of a group of entities is in some sense better than individual ones. Pioneered by Galton \cite{galton}, this effect was in its simpler incarnation a direct consequence of the law of large numbers. Evidence of collective intelligence spans today social systems in different species \cite{bonabeau, human, insects} and activities ranging from optimal estimation \cite{bonabeau}, navigation \cite{navigation}, or sensing \cite{fish} to cite a few. In this work we focus on the phenomenon of collective musical performance. We are not interested in choral performances but on self-organized 'crowd performances' that take place in popular music concerts \cite{youtube}, sport events (e.g. in football stadiums) or other social events, or simply within groups of people that join together to perform a song or melody. Our contention is that whereas the average individual is not necessarily a gifted performer and does not particularly sing {\it in tune} (i.e. individual musical performances are typically of poor quality), when a large group of these imperfect singers perform at unison the resulting collective signal is surprisingly {\it tuned}. As a consequence, crowd performance is enhanced as compared to individual ones and is thus perceived as a choral one. Whereas some research suggests that individuals improve while performing at unison \cite{children} -pointing that imitation might be underpinning this phenomenon-, here we show that imitation, while clearly boosting this effect, is not itself required for the enhancement to occur in the first place. We present a toy model that supports this claim and that provides a simple explanation for the origin of this collective phenomenon.\\

\noindent {{\bf Perceived pitch. }It should be stressed that in order to assess whether a crowd or an individual has a good intonation, the listener evaluates such intonation on the basis of the tone he perceives, or more accurately, based on the perceived pitch. Now, pitch is indeed a perceptual (subjective) property of sound \cite{Chialvo3}, a psychoacoustic phenomenon more similar to a sensation synthesized by the brain than an objective reality. The perceived pitch indeed coincides with the frequency in the case where sound is formed by a pure tone (sinusoidal wave). In the more realistic case of a complex tone -a sound composed by several tones-, the perceived pitch is a vaguely-defined concept which has been the source of debate and research since the 19th century. To substantiate this statement, we can refer to the so called missing fundamental illusion, originally discovered by Seebeck in 1841 and observed experimentally by de Boer \cite{Boer} and Shouten et al. \cite{Schouten}, which dictates that when several harmonics are played together, in some circumstances the perceived pitch does not correspond to any of the frequencies at which the air is vibrating but indeed corresponds to a frequency which is not physically present.\\
In our toy model, the melody to be played consists of a pure tone with frequency $T$, which the crowd interprets at unison. As individuals (or agents, from now on) don't usually have a perfect pitch, the collective output produced by the crowd will be a inharmonic complex tone that develops out of the mixture of each agent's contribution. For simplicity we will assume that each agent contributes with a pure tone -i.e. a sinusoidal wave with a single frequency-.
\noindent Now, consider two tones with frequency $f_1$ and $f_2$ and similar amplitudes, played simultaneously. What is the pitch of this complex tone?  If $f_1$ and $f_2$ are sufficiently close, then the pitch is somewhere close to $(f_1+f_2)/2$ and is accompanied with a beating at $|f_1-f_2|$. As their difference increases the beating disappears, and for sufficiently different frequencies one can indeed perceive both frequencies. Furthermore, if $f_2=pf_1$ for some integer $p$, then the pitch of the complex tone is just $f_1$, coinciding with the fundamental frequency that corresponds to the greatest common divisor between both frequencies, GCD$(f_1,f_2)$. Remarkably, if two tones $f_1$ and $f_2$ are harmonically related with a third one $f_0$ such that $f_1=pf_0$ and $f_2=p'f_0$ with $p,p'>1$, then the perceived pitch reduces again to the fundamental frequency $f_0$, which in this case is not physically present. Despite the absence of energy at $f_0$ (there is no actual source of air vibration at that frequency) as a result of constructive interferences the missing fundamental frequency emerges as the perceived pitch. Schouten called this missing fundamental the residue pitch \cite{Schouten, heller}.} For a complex tone with $N>2$ partials, the story is far more intricate. {Let us consider initially the case where all frequencies are harmonically related}. If we superpose frequencies $f,2f,3f,...$ (that is, a fundamental $f$ and a few higher harmonics) with more or less the same amplitude, then the resulting pitch will indeed be $f$, coinciding again with the fundamental frequency.  If we now remove $f_0$ and only superimpose $2f_0,3f_0,4f_0,\dots,Nf_0$ again GCD$(2f_0,3f_0,\dots,Nf_0)=f_0$:  we will still perceive $f$ which in this case is, as before, a missing partial, the residue pitch of Schouten. This happens as in the range 20-2000Hz, the ear has the ability to fuse harmonically-related frequencies into a single entity with a fundamental frequency, even in the case where such fundamental frequency is missing. 
Now, in general the fundamental frequency (either physically present or missing) does not {\it necessarily} correspond to the {\it perceived pitch}, i.e., to the effective frequency perceived by an external agent who is listening to the collective output: pitch is a psychoacoustic phenomenon far more complex than basic frequency superposition. 
This fact becomes evident when we combine partials which are not harmonically related. Consider the mixture of frequencies at $120$, $220$,
$320$, $420$, $520$, and $620$ Hz in equal measure. The GCD of the mixture is $20$ Hz (a frequency which is indeed barely audible), however the perceived pitch coincides in this case with a mysterious frequency located at $104.6$ Hz \cite{heller}.\\
\begin{figure*}
\centering
\includegraphics[width=0.86\columnwidth]{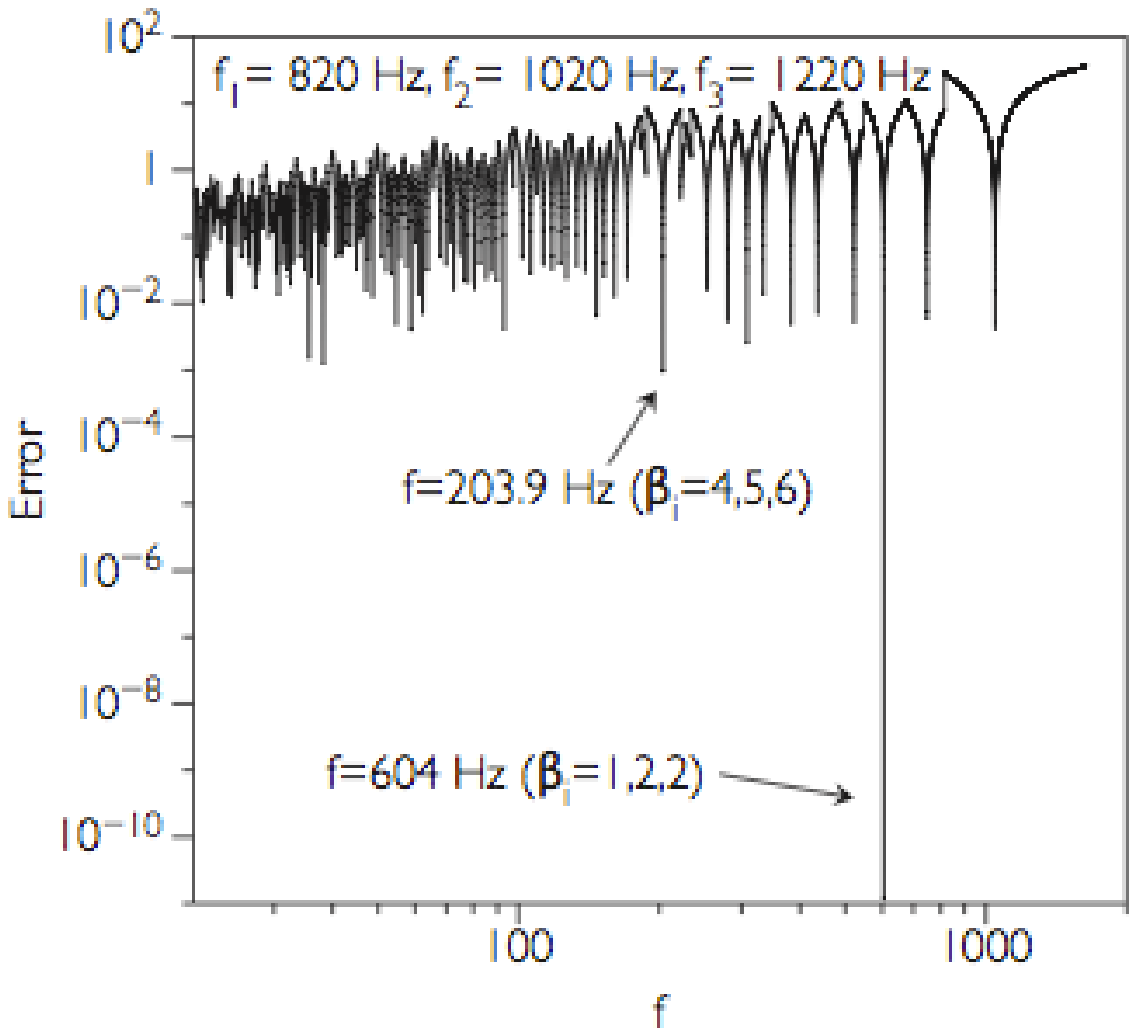}
\includegraphics[width=0.9\columnwidth]{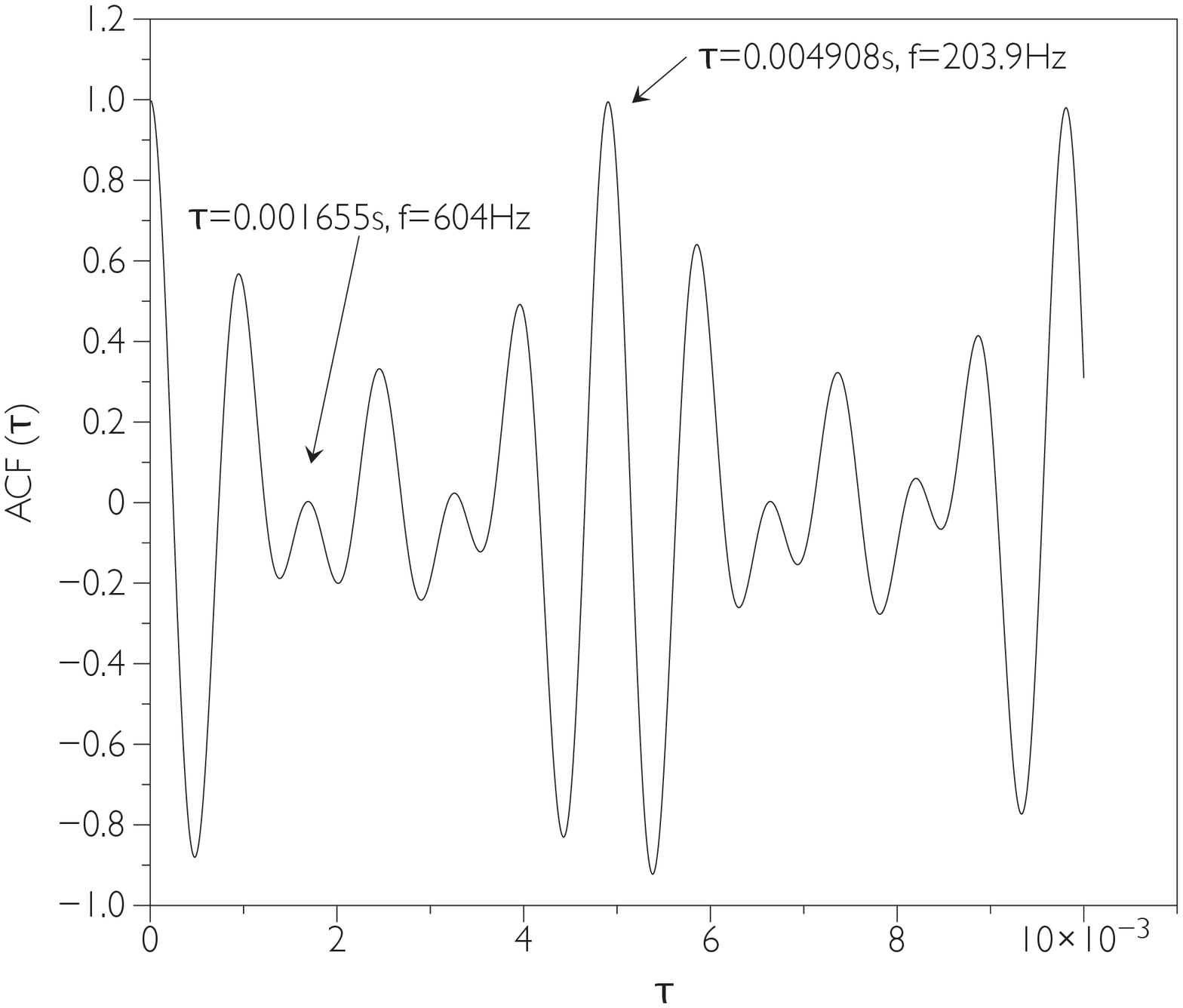}
\caption{(Left panel) Numerical evaluation of the roots of equation \ref{basic} for a complex tone of three partials with $f_1=820$ Hz, $f_2=1020$ Hz, $f_3=1220$ Hz. Each solution -denoted as a frequency with low numerical error- corresponds to a local peak in the autocorrelation function (and its harmonics). Two possible solutions with different $\beta_i$ are depicted. The perceived pitch is indeed $\bar f=203.9Hz$ \cite{heller, Pierce}, which corresponds to $\beta_i=4,5,6$.
(Right panel) Autocorrelation function of the same complex tone, where one can appreciate that the perceived pitch is indeed associated with the first non-trivial 'large peak', whereas other peaks that take place sooner are not strong enough to develop into the perceived pitch.}
\label{Fig1}
\end{figure*}

\noindent There are essentially two theories (or groups of theories) on pitch perception \cite{heller, Greenberg, Cariani, Tramo}, namely those focusing on spatial separation of partials in the ear (Fourier decomposition theories pioneered by Ohm and Helmholtz) and those that focus on the temporal separation, pioneered by Licklider \cite{Licklider}. These are not necessarily mutually exclusive, and both might have their range of validity. {The problem of the missing fundamental or residue pitch has also been extensively addressed, and a variety of mechanisms operating underneath have been proposed, ranging from delay lines \cite{Licklider}, integration circuits \cite{Langner}, timing nets \cite{Cariani} or neural networks \cite{Cohen} to cite some. Recent works have linked this very same problem to a mechanism entitled ghost stochastic resonance \cite{Chialvo1, Chialvo3}, by which linear interference of tones is boosted nonlinearly by using noisy thresholds, thus providing a minimal and biologically plausible mechanism by which the strongest resonance is the one that enhances the missing fundamental \cite{Chialvo2}.\\
Here we adhere to the perspective where pitch is associated with something more fundamental than the presence or absence of a particular partial: its tendency to repeat itself at given
intervals. The extension of this concept to signals that are not strictly
periodic is the autocorrelation function. 
There is evidence that temporal theories -i.e., based on autocorrelation- indeed apply to most of what happens below 5000 Hz, which is where all music belongs. Spatial (Fourier) theories on the other hand play a prominent role above 5000 Hz. As a matter of fact, our neural processes cannot keep track of time intervals shorter than about 0.0002 seconds, or 5000 Hz, so it is reasonable that we lose the precision of timing above that frequency and switch over to a place mechanism for detecting frequency -the region of maximum excitation of the basilar membrane- above 5000 Hz \cite{heller}.
 Interestingly, in the example above with a complex inharmonic mixture of six frequencies, the first non-trivial peak of the signal's autocorrelation function occurs precisely at $t=0.009565 s$ which is related to a periodic repetition at $f=104.6$ Hz, coinciding with the mysterious perceived pitch.}\\ 
To be more precise, let us consider a complex tone of $N$ sinusoidal partials with frequency $f_i$ and amplitude $a_i$. The resulting signal is $s(t)=\sum_{i=1}^N a_i \sin(2\pi f_i t)$, and let us denote by $\bar f$ the perceived pitch of this mixture. Then $\bar f$ coincides with $1/\tau_M$ where $\tau_M$ is the time position of the earliest tall peak in the autocorrelation function $C(\tau)=\langle s(t)s(t+\tau)\rangle_t$. This is an extremum thus $dC/d\tau|_{\tau_M}=0$. Moroever, as the product $\sin(t)\sin (t+\tau)$ is maximized for $\tau$ being a multiple of $2\pi$, it is easy to see that for any local peak at $\tau_M$ of the autocorrelation function one has $\sin(2\pi f_i \tau_M)\approx 2\pi (f_i\tau_M-\beta_i)$, for some integer $\beta_i$. Putting all these conditions together, according to Heller \cite{heller} the peaks of the autocorrelation function fulfil the following self-consistent equation
\begin{equation}
1/\tau_M= \bar f \approx \frac{\sum_{i=1}^N a_i^2f_i^2} {\sum_{i=1}^Na_i^2\beta_i f_i},
\label{basic}
\end{equation}
where for $i=1,\dots,N$, $\beta_i \in \mathbb{Z}$ is the nearest integer to $f_i/\bar{f}$. This formula was first derived in the context of molecular spectroscopy to account for the so-called missing mode effect (MIME) \cite{MIME1, MIME2, MIME3}. It is important to highlight that $\bar{f}$ is not just a convoluted average of each frequency \cite{MIME2}, but in some sense is an emergent quantity out of the combination of partials, much like 
in the luminescence spectra of complex molecules some regularly spaced vibronic progressions emerge even if they don't correspond 
to any ground-state normal mode of vibration (or average) of the molecule \cite{MIME3}. {In other words, despite the fact that pitch is a psychoacoustic phenomenon, it is still quantitative \cite{heller}, as it can be associated to objective features such as the autocorrelation function of the signal.}\\

\noindent We solve eq.\ref{basic} numerically and assume that
$f_{\text{app}}$ is a good approximation to $\bar f$ if
$$\bigg(f_{\text{app}}- \frac{\sum_{i=1}^N a_i^2f_i^2} {\sum_{i=1}^Na_i^2\hat{\beta}_i f_i}\bigg)\cdot 100/f_{\text{app}} < \epsilon,$$
where $\hat{\beta}_i$ is the nearest integer to $f_i/f_{\text{app}}$. As a rule of thumb, we set $\epsilon = 10^{-2}$, which means that $f_{\text{app}}$ satisfies eq. \ref{basic} self-consistently with an error which is less $0.01\%$ of the frequency $f_{\text{app}}$.\\
\begin{figure*}
\centering
\includegraphics[width=0.9\columnwidth]{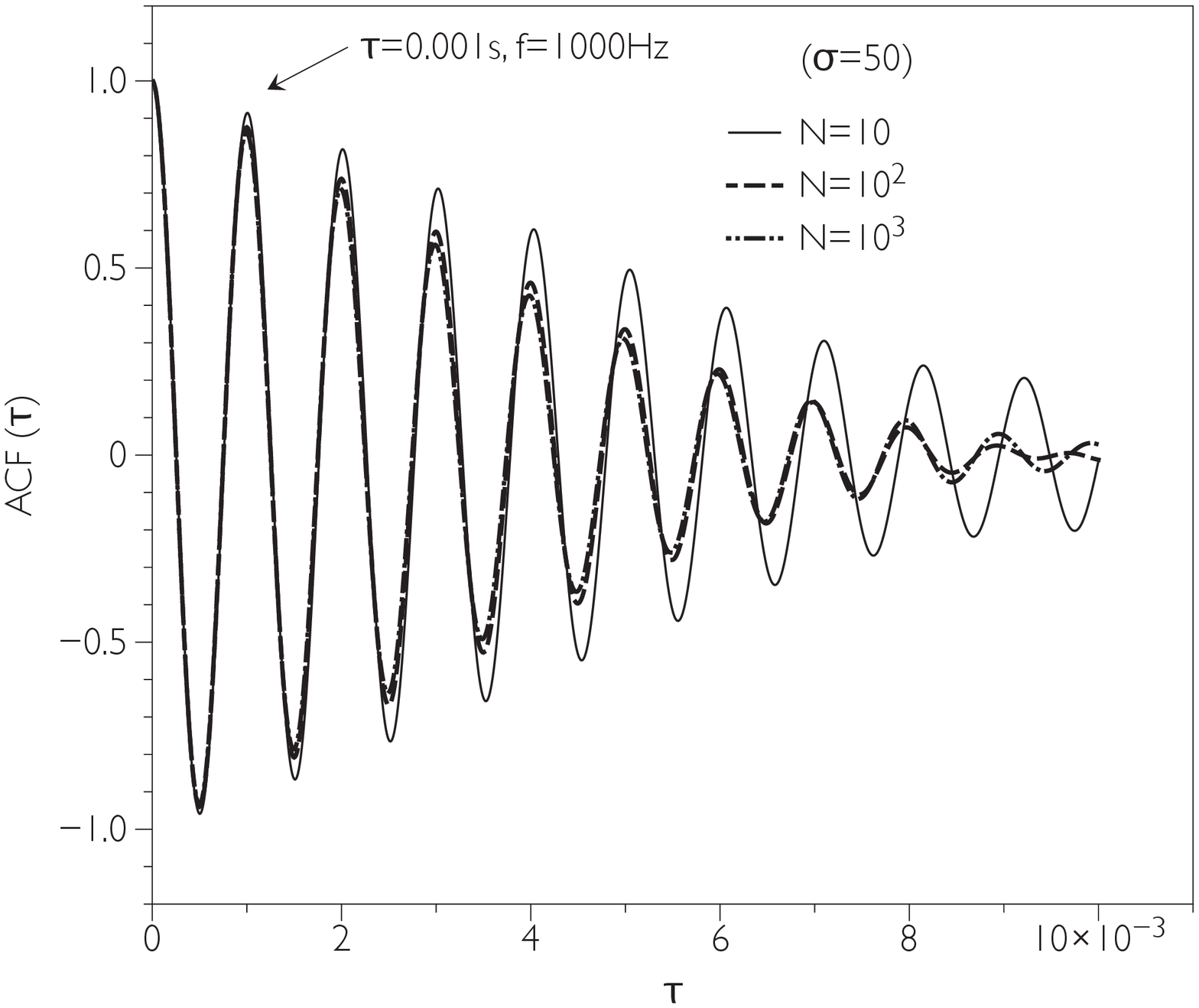}
\includegraphics[width=0.9\columnwidth]{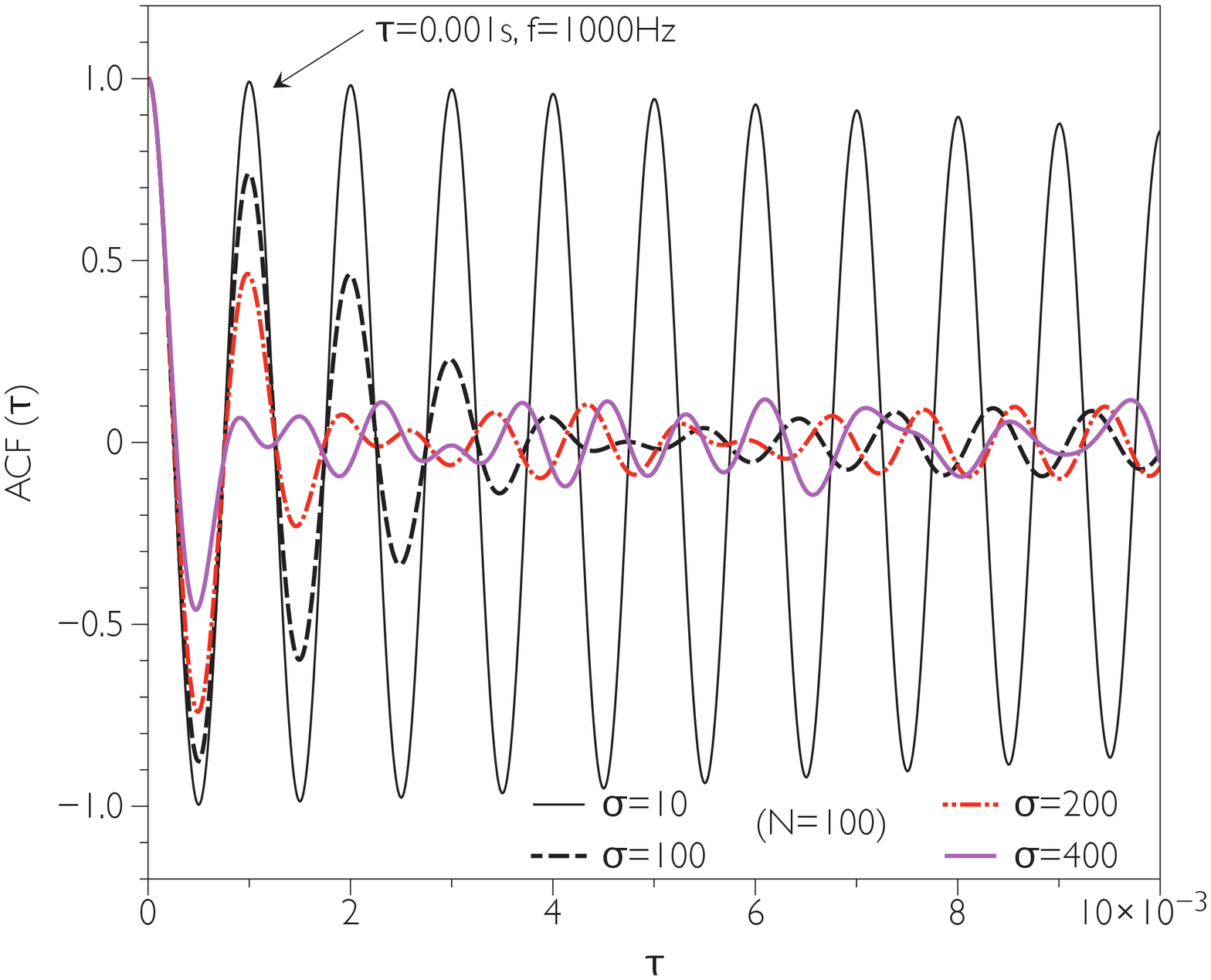}
\caption{(Left panel) Autocorrelation function of a complex tone formed by $N$ frequencies $f_i \sim {\cal N}(1000,50)$, for $N=10$ (solid line), $N=10^2$ (dashed line) and $N=10^3$ (dashed dotted line). In every case the first non-trivial large peak in the autocorrelation function lies at $\tau=0.001$ seconds, yielding a perceived pitch $\bar{f}=1000$ Hz. Interestingly, other peaks in the autocorrelation function (associated with other solutions of eq. \ref{basic}) vanish as $N$ increases with a  where one can appreciate that the perceived pitch is indeed associated with the first non-trivial 'large peak', whereas other peaks that take place sooner are not strong enough to develop into the perceived pitch. (Right panel) Autocorrelation function of a complex tone formed by $N=100$ frequencies $f_i \sim {\cal N}(1000,\sigma^2)$, for increasing values of $\sigma^2$. The perceived pitch converges to $\bar{f}=1000$Hz for a rather large range of values $\sigma^2$, after which the complex tone does not have a clear perceived pitch.}
\label{fig_2}
\end{figure*}
\noindent Now, it is easy to observe that that eq.\ref{basic} is multivalued: at least it admits as solutions the perceived pitch and its infinitely many subharmonics. Indeed, for equal frequencies $f_i=k \ \forall i$, the perceived pitch is trivially $\bar f=k$ and this is a solution of eq.\ref{basic} for $\beta_i=1 \ \forall i$, but so are subharmonics $k/p$, for $p\in \mathbb{N}^+$ and $\beta_i=p \ \forall i$. Also, for two close enough frequencies $f_1\approx f_2$, then $(f_1+f_2)/2$ is an approximate solution which indeed corresponds to the perceived pitch. Now, eq.\ref{basic} captures the location of peaks in the autocorrelation function, but unfortunately not their height. Consider for instance the complex tone formed by partials of equal amplitude at frequencies 820, 1020, and 1220 Hz. The GCD is 20 Hz, right at the threshold of hearing, and seems an unlikely perceptual result of
combining these much higher frequencies. Pierce \cite{Pierce} cites this case as an interesting example and
reports that the perceived pitch is 204 Hz. A possible solution can be found for $\beta_1=1, \beta_2=\beta_3=2$, for which setting $a_i=1$, we get $\bar f=604$ Hz. According to the left panel of fig.\ref{Fig1}, this seems indeed the solution with minimal numerical error. The solution with second minimal error corresponds to a higher combination $\beta_1 = 4$, $\beta_2 =
5$, $\beta_3 = 6$ for which $\bar f \approx 203.9$ Hz. However, it is this latter candidate that coincides with the empirical value found by Pierce. If we look at the autocorrelation function of the complex tone (right panel of the same figure), we indeed discover peaks at $1/604=0.001655$ and $1/203.9=0.004908$ seconds (among others), however the latter is the sharpest peak and hence constitutes the perceived pitch.\\
All in all, the systematic computation of the perceived pitch is not straightforward. Heller \cite{heller} speaks about three criteria to determine what peak corresponds to the perceived pitch of a complex tone: (i) the sooner in the autocorrelation function (sooner times corresponds to larger frequencies), (ii) the larger the autocorrelation of the peak, and (iii) the sharper the peak. However looking at the solutions of eq.\ref{basic} we are only able to discern criterion (i), therefore in what follows we will focus on the autocorrelation function to discern the perceived pitch from the set of solutions of eq.\ref{basic}.\\

\noindent {\bf Basic model. } We return now to the toy model briefly discussed above. Consider $N$ agents aiming to sing at unison a given frequency $T$. 
We assume that all agents sing pure tones (i.e. sinusoids of frequency $f_i$) at approximately the same amplitude ($a_i=K \ \forall i$ for some $K\in \mathbb{R}^+$) and model the imperfection of each agent as an independent Gaussian deviation. That is, $\forall i=1,\dots,N$ the frequency $f_i=T + \xi$, where $\xi \sim {\cal N}(0,\sigma^2)$. The standard deviation $\sigma$ therefore tunes the diversity of imperfections. Note that trivially, $\lim_{N\to \infty} \text{GCD}(f_1\dots f_N)=0$. Is there a perceived pitch for this complex tone? 
Applying eq.\ref{basic} in this case, one finds a frequency $$\bar f \approx \frac{\sum_{i=1}^N (T+\xi_i)^2}{\sum_{i=1}^N \beta_i(T+\xi_i)}=\frac{\sum_{i=1}^N T^2+ \sum_{i=1}^N\xi_i^2 + 2\sum_{i=1}^N T\xi_i}{\sum_{i=1}^N\beta_iT+\sum_{i=1}^N\beta_i\xi_i}.$$
To prove that the crowd sings better than each individual in a nontrivial way, we need to (i) find that $\bar{f}\approx T$ is a solution to the latter equation that (ii) corresponds to an early tall peak in the autocorrelation function of the signal and that (iii) this holds for a range of values of $\sigma$. The criterion (iii) is required as if the phenomenon only holded for very small $\sigma$, one could argue that in practice the perceived pitch would be harmonically fusing barely audible deviations from the correct pitch. In the contrary, if $\sigma$ is large enough such that every random sample is almost surely out of tune then the emergence of a tuned perceived pitch would be a genuine emergent phenomenon.\\
First, as eq.\ref{basic} is multivalued for simplicity we focus in the solution associated to $\beta_i=1\  \forall i$. In this case, trivially $\sum_{i=1}^N T^2= NT^2$ and $\sum_{i=1}^N\beta_i T=NT$. According to the central limit theorem, the sum of $N$ Gaussian random variables ${\cal N}(0,\sigma^2)$ variables is a Gaussian random variable ${\cal N}(0,N\sigma^2)$. Thus  for $N\gg 1$ we can use expected values such that $\sum_{i=1}^N \beta_i\xi_i\to 0$ and $\sum_{i=1}^N \xi_i^2 \to N\langle \xi^2 \rangle=N\sigma^2$ (alternatively, the sum of $N$ squared standard Gaussian random variables is a random variable which is distributed as a $\chi^2$ distribution with mean $N$, so if the original Gaussian variables are not standard but have variance $\sigma^2$, then the mean of the rescaled $\chi^2$ distribution is $N\sigma^2$). Altogether, the solution to eq.\ref{basic} associated to $\beta_i=1\  \forall i$ is 
\begin{equation}
\bar f \approx \frac{NT^2 + N\sigma^2}{NT}=T+\sigma^2/T
\label{baseline}
\end{equation}    
Provided that extremely large deviations from the correct tone are not abundant among individual performance (so that $\sigma \ll T$ is a good approximation) then the second term in the latter solution is $\ll T$ and then at leading order $\bar f \approx T$. Now, to evaluate whether this frequency indeed corresponds to the earliest tall peak in the autocorrelation funcion, we have tested this prediction numerically in figure \ref{fig_2}.  While human hearing ranges from 20 to 20000 Hz, the greater sensitivity is known to lie within 200 and 2000 Hz. We therefore discard solution frequencies under 100 Hz (that is, times larger than $10^{-2}$ seconds) as they will only contribute to perceived background noise. In the left panel of figure \ref{fig_2} we plot the autocorrelation function of a complex tone made by $N$ sinusoids with equal amplitude and frequencies $f_i \sim {\cal N}(T,50)$, for $T=1000$ Hz. We can observe that as the number of agents $N$ increases, the frequency $\bar{f}=T$ indeed emerges as the clear perceived pitch (numerical evaluation of the solutions of eq.\ref{basic} are plotted in an appendix figure). In the right panel of the same figure we explore the effect of increasing $\sigma$ in the shape of the autocorrelation function for $N=100$ frequencies (solutions of eq.\ref{basic} in this case are again summarized in an appendix). As expected, the frequency $\bar{f}=1000$ Hz coincides with the perceived pitch for a reasonably large range of values of $\sigma$. For $\sigma=10$ and $100$ the peak is clearly visible although it decreases as $\sigma$ increases. Note that the just noticeable difference (which quantifies the threshold at which a change in pitch is perceived) depends on the frequency and for $1000$ Hz this is smaller than 10 Hz. This means that for $\sigma=100$ most of the individual agents will be effectively out of tune, however the perceived pitch of the aggregate will still emerge as being in tune. Accordingly, the emergent pitch is robust even if each agent is not particularly gifted, musically speaking. For even larger values ($\sigma=400$) the spectrum approaches a flat shape, the peak has faded away and no clear pitch emerges accordingly. All in all, we can conclude that a crowd indeed sings better as a whole than each individual separately, even if no synchronization takes place among individuals.\\

\noindent {\bf Introducing short-range interactions. }
Remarkably, our basic model suggests that it is not required that each agent interacts for the perceived pitch to collective emerge as the 'correct' intonation. It is however true that in realistic cases individuals that sing in groups tend to tune up with their surrounding, if in their close neighborhood there is at least some other person with better intonation \cite{children}. While intonation and imitation capacities are definitely heterogeneous across people,  in what follows we show that under general conditions this imitation process effectively reduces the value of $\sigma$ in eq.\ref{baseline}.\\ 
\noindent To explore the effect of imitation we propose the following toy model: we {model a crowd as a set of agents located in} the vertices of a two-dimensional lattice. Each agent $i$ vibrates (sings) at a given frequency $f_i^{(t)}$ which can now be dynamically updated. Initially we again assign $f_i^{(0)}=T + \xi$, for Gaussian i.i.d. random variables $\xi \sim {\cal N}(0,\sigma^2)$. Then, at each simulation time $t$ each agent $i$ updates its frequency according to the following rules: {if any neighbor is singing more {\it in tune} than $i$ (modeled by the fact that in the Von Neumann neighborhood of $i$ we find $|f_i^{(t)}-T| > \min \{|f_j^{(t)}-T|\}_{\text{nn}}$), then agent $i$ updates his frequency after a process of imperfect imitation with the agent with better intonation, such that his updated frequency reads
$f_i^{(t+1)}=f_i^{(t)} + C(i)[{\text{M}} - f_i^{(t)}]$,
where ${\text{M}}$ is the frequency} to be imitated and $C(i)\in [0,1]$ is a real number that describes the fitness of agent $i$ to imitate or tune up. Intuitively, an agent with good imitation skills will initially perform close to $T$, so for simplicity we define $C(i)=1-\frac{|f_i^{(0)}-T|}{T}$. {Accordingly, for initial performances close to $T$, $C$ will be close to 1, and conversely for bad initial performances the agent will have a low capacity, $C$ close to zero.}\\
 {Parallel iteration of this updating process models the adaptation and imitation of agents over time}. The relevant observable of the system is again the perceived pitch $\bar f(t)$ which is now a function of time and will change as the frequencies variance $\sigma^2(t)$ is modified.
\begin{figure}
\centering
\includegraphics[width=0.8\columnwidth]{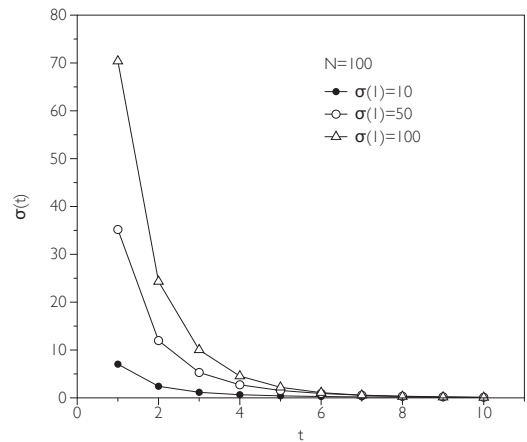}
\caption{Numerical evaluation of standard deviation of the distribution of frequencies of the lattice model, as interactions take place over time. Already after one simulation step, the effective standard deviation considerably decreases, what in turn implies that the virtual pitch approximates its leading order $\bar f \approx T$}
\label{fig_4}
\end{figure}
If imitation is null ($C(i)=0$) then this model reduces to the non-interacting case above. At the other extreme, if every agent has perfect imitation skills ($C(i)=1$) then there is an absorbing state where all the agents end up vibrating at the same characteristic frequency $f_i^*$ that corresponds to the one for which $|f_i^{(0)}-T|$ is minimized. That is to say, amongst the initial values of the partials, the one closest to $T$ percolates and emerges as a consensus. In this ideal situation, it is easy to see that as $N \to \infty$, $\bar f(\infty) \to T$. In the more realistic case where $C(i) \in (0,1)$, the absorbing state will be such that $\xi_i^{(\infty)}$ will not just have one value but several (corresponding to several degrees of intonation). However what is straightforwardly guaranteed is that the variance of the frequencies distribution $\sigma^2$ will decrease over time with respect to the initial condition (non-interacting case). That is to say, the second term in eq.\ref{baseline} will get necessarily by monotonically decreasing over time, boosting even further the collective intonation effect. These tendencies are confirmed by numerical simulations in figure \ref{fig_4}.\\
As a final comment, note that in the event that the crowd is the audience of a concert which follows the band's lead singer (i.e. the system is coupled to an external 'pitch field'), then the imitation process directly takes place with the singer, instead of locally. This mechanism trivially uncouples the system and reduces the problem to the original non-interacting case, albeit with a new $\sigma$ which is much smaller than in the original case.\\

\noindent To conclude, we have given a simple explanation for the emergence of collective intonation in crowds that sing at unison. Within reasonable limits, regardless the intonation of each singer the collective tone will be perceived as to be in tune. This collective effect is further boosted if one allows individuals to adjust their frequency by any degree of imperfect imitation with his neighbors, although, remarkably, this additional mechanism is not required for the collective effect to emerge in the first place. Furthermore, this result does not require subjects to follow any leader, and emerges in a self-organized way due to the psychoacoustic properties of the perceived pitch.\\

\acknowledgments{The author thanks Andrew Berdahl for fruitful discussions and encouragement and Dante Chialvo for showing the relation with ghost stochastic resonance.}



\section{Appendix}

\newpage

\begin{figure}[h]
\centering
\includegraphics[width=0.7\columnwidth]{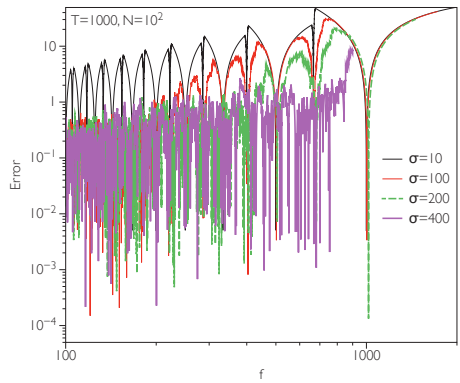}
\caption{Numerical evaluation eq.\ref{basic} for $N=10^2$ frequencies $f_i\sim {\cal N}(T=1000,\sigma^2)$. For $\sigma \ll T$, there are few solutions that consist of $T$ and its subharmonics. As $\sigma$ increases, other solutions start to appear, and $\bar f=T$ eventually disappears.}
\label{fig_appendix2}
\end{figure}

\begin{figure}[h]
\centering
\includegraphics[width=0.75\columnwidth]{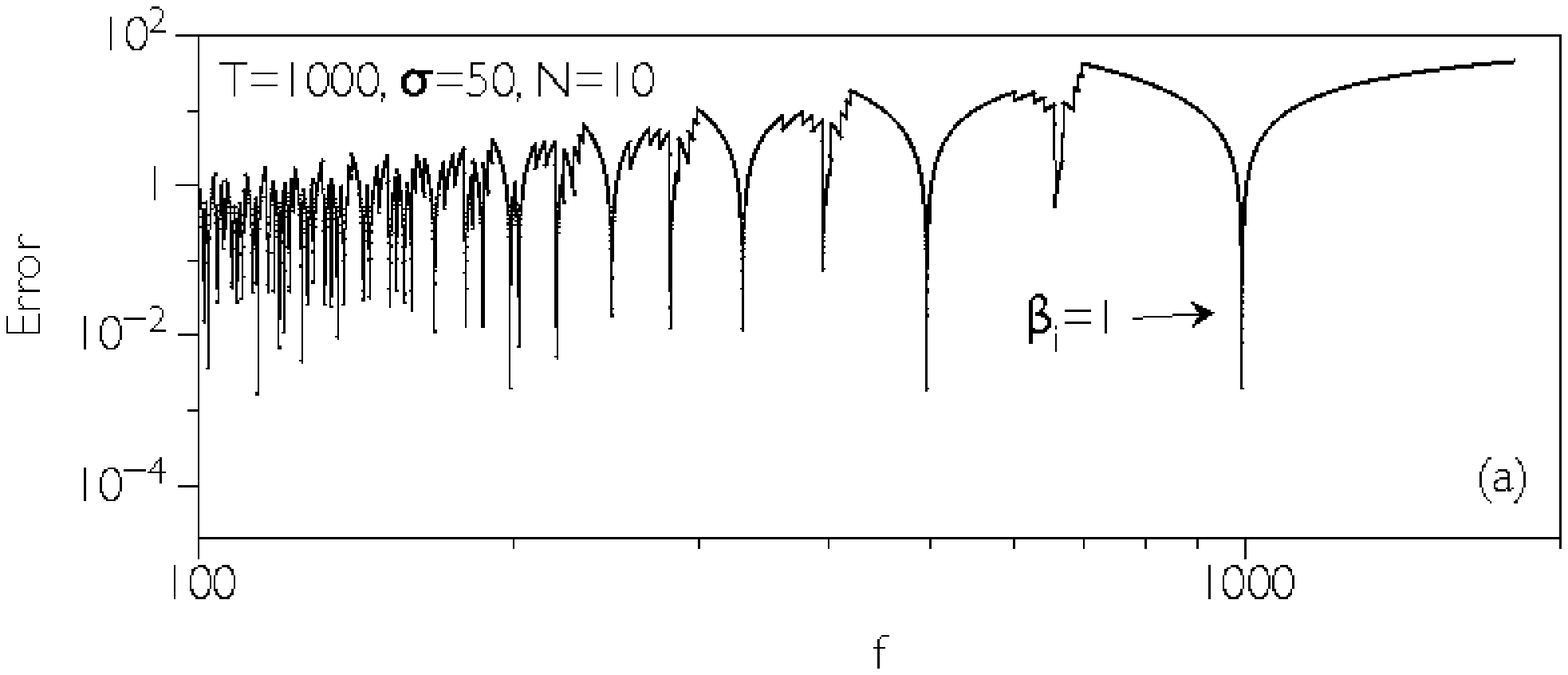}
\includegraphics[width=0.75\columnwidth]{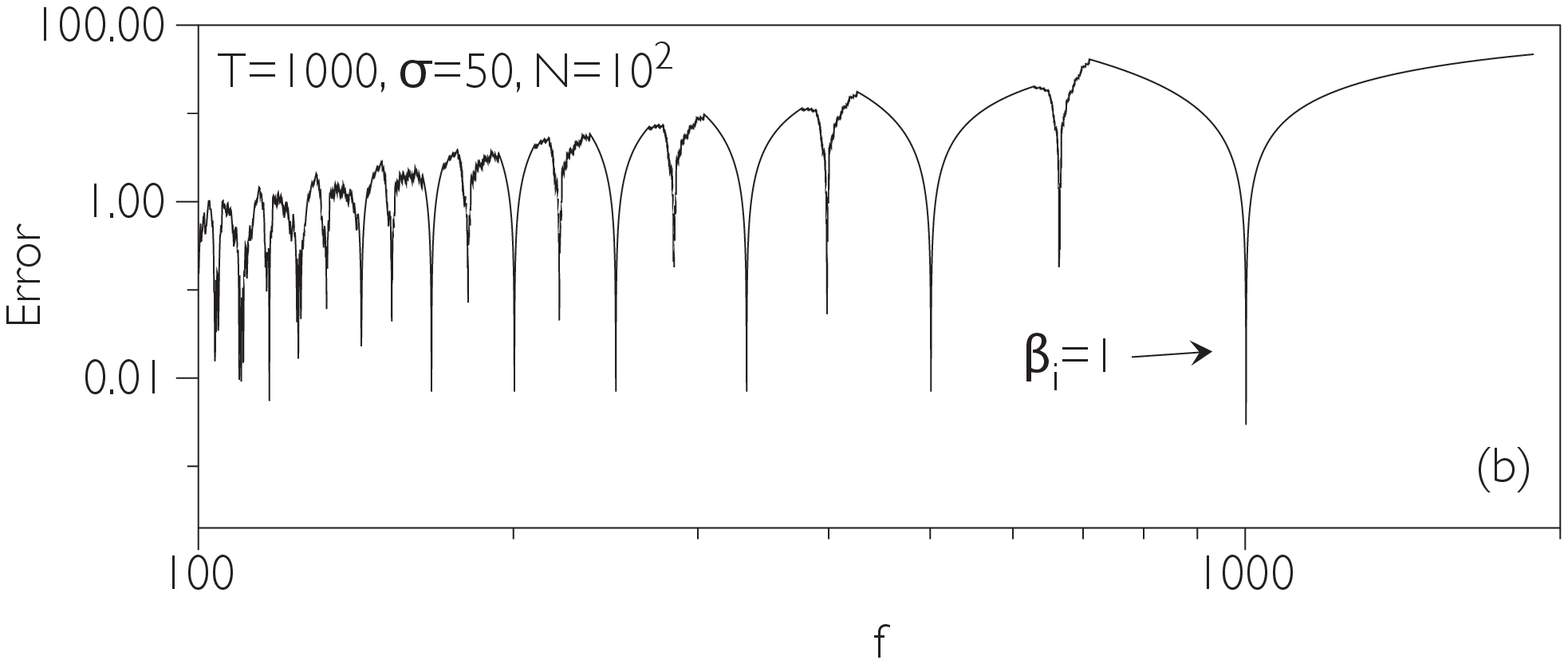}
\includegraphics[width=0.75\columnwidth]{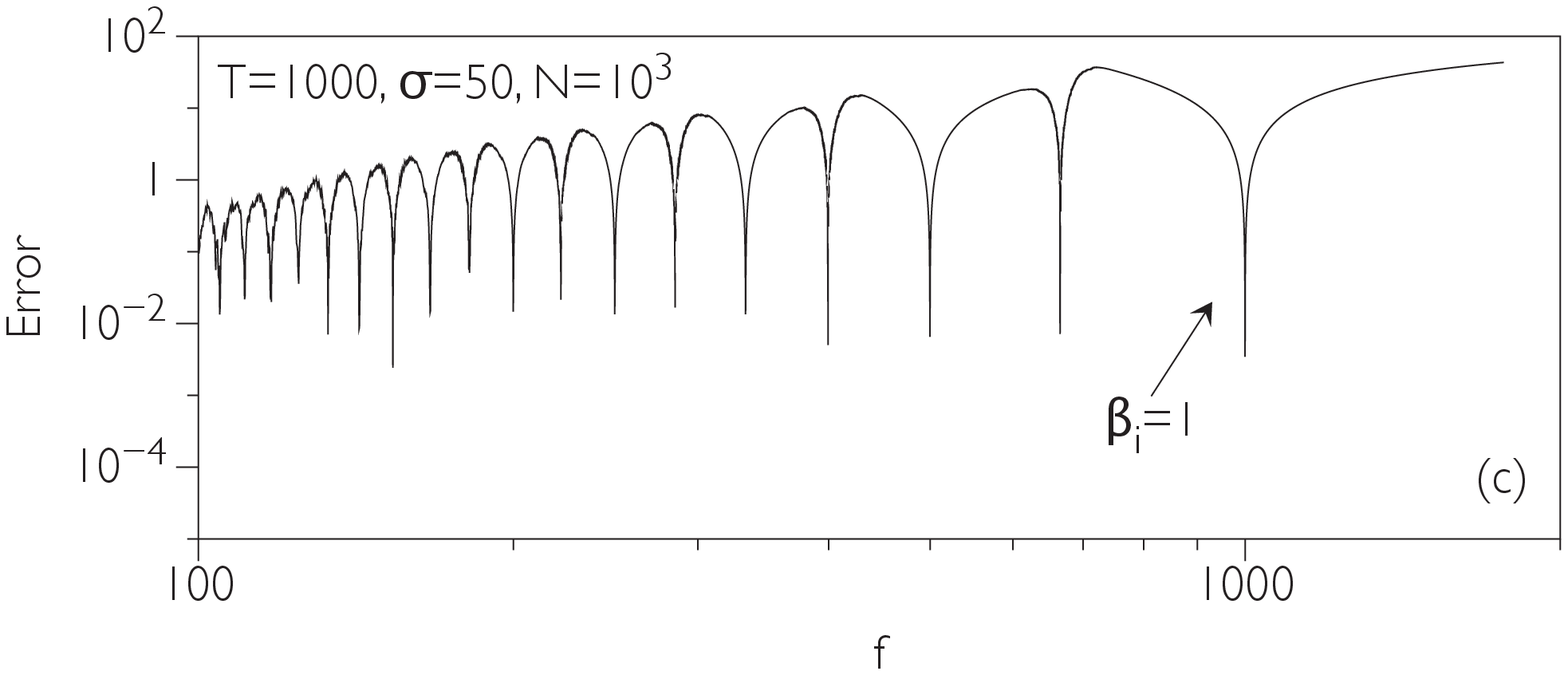}
\caption{Numerical evaluation of the roots of equation \ref{basic}, where frequencies $f_i\sim {\cal N}(1000,50)$ for $N=10, 10^2$ and $10^3$  respectively. As $N$ increases, just a few frequencies (and subharmonics) emerge as the numerical solutions.}
\label{fig_appendix1}
\end{figure}

\end{document}